\documentclass[utf8]{FrontiersinHarvard} % for articles in journals using the Harvard Referencing Style (Author-Date), for Frontiers Reference Styles by Journal: https://zendesk.frontiersin.org/hc/en-us/articles/360017860337-Frontiers-Reference-Styles-by-Journal
\usepackage{url,hyperref,lineno,microtype,subcaption}
\usepackage[onehalfspacing]{setspace}
\usepackage{lmodern}
\usepackage{hyperref}
\usepackage{amsmath}

\def\keyFont{\fontsize{8}{11}\helveticabold }
\def\firstAuthorLast{Kunpeng Mao {et~al.}} %use et al only if is more than 1 author
\def\Authors{Kunpeng Mao\,$^{1,+}$, Ruoyu Li\,$^{2,+}$, Junlong Cheng\,$^{2}$, Danmei Huang\,$^{1}$, Zhiping Song\,$^{1,*}$ and ZeKui Liu\,$^{1,*}$}
\def\Address{$^{1}$Chongqing City Management College, Chongqing, China \\
$^{2}$College of Computer Science, Sichuan University, Chengdu, China \\
$^{+}$These authors contributed equally to this work. }
\def\corrAuthor{Zhiping Song and ZeKui Liu}
\def\corrEmail{2452671073@qq.com and 704603595@qq.com}
\usepackage{multirow}
\begin{document}
\onecolumn
\firstpage{1}

\title[Running Title]{PL-Net: Progressive Learning Network for Medical Image Segmentation} 

\author[\firstAuthorLast ]{\Authors} %This field will be automatically populated
\address{} %This field will be automatically populated
\correspondance{} %This field will be automatically populated

\extraAuth{}% If there are more than 1 corresponding author, comment this line and uncomment the next one.
%\extraAuth{corresponding Author2 \\ Laboratory X2, Institute X2, Department X2, Organization X2, Street X2, City X2 , State XX2 (only USA, Canada and Australia), Zip Code2, X2 Country X2, email2@uni2.edu}

\maketitle

\begin{abstract}
In recent years, deep convolutional neural network-based segmentation methods have achieved state-of-the-art performance for many medical analysis tasks. However, most of these approaches rely on optimizing the U-Net structure or adding new functional modules, which overlooks the complementation and fusion of coarse-grained and fine-grained semantic information. To address these issues, we propose a 2D medical image segmentation framework called Progressive Learning Network (PL-Net), which comprises Internal Progressive Learning (IPL) and External Progressive Learning (EPL). PL-Net offers the following advantages: (1) IPL divides feature extraction into two steps, allowing for the mixing of different size receptive fields and capturing semantic information from coarse to fine granularity without introducing additional parameters; (2) EPL divides the training process into two stages to optimize parameters and facilitate the fusion of coarse-grained information in the first stage and fine-grained information in the second stage. We conducted comprehensive evaluations of our proposed method on five medical image segmentation datasets, and the experimental results demonstrate that PL-Net achieves competitive segmentation performance. It is worth noting that PL-Net does not introduce any additional learnable parameters compared to other U-Net variants.
%%% Leave the Abstract empty if your article does not require one, please see the Summary Table for full details.

\tiny
 \keyFont{ \section{Keywords:} Progressive Learning, Coarse-grained to Fine-grained Semantic Information, Complementation and Fusion, Medical Image Segmentation} %All article types: you may provide up to 8 keywords; at least 5 are mandatory.
\end{abstract}

\section{Introduction}

Medical image segmentation is a technique used to extract regions of interest for quantitative and qualitative analysis. For example, it can be used for cell segmentation in electron microscopy (EM) recordings \citep{ref2}, melanoma segmentation in dermoscopy images \citep{ref3,ref4}, thyroid nodule segmentation in ultrasound images , and heart segmentation in MRI images \citep{ref5}. Traditionally, medical image segmentation methods relied on manually designed features to generate segmentation results \citep{ref6,ref7}. However, this approach requires distinct feature designs for various applications. Furthermore, the large variety of medical image modalities makes it difficult or impossible to transfer a specific type of feature design method to different image types. Therefore, the development of a universal feature extraction technique is crucial in the field of medical image analysis.

The emergence of deep learning technology has revolutionized medical image segmentation by overcoming the limitations of traditional manual feature extraction methods. Convolutional neural networks (CNN) based automatic feature learning algorithms, such as the fully convolutional network (FCN) proposed by Shelhamer et al. \citep{ref8} and the U-Net framework for biomedical image segmentation proposed by Ronneberger et al., \citep{ref2} have shown promising results. The FCN model structure is designed to be end-to-end, which eliminates the need for manual feature extraction and image post-processing steps. On the other hand, the U-Net framework's encoder-decoder-skip connection network structure has shown good results on medical image segmentation datasets with small amounts of data. %To further improve performance in different tasks, researchers have developed many variants of U-Net. These variants either add new functional modules to U-Net or optimize its structure. Therefore, the application of deep learning technology in medical image segmentation has opened avenues for more accurate and efficient diagnoses in the medical field.

%为了进一步提高模型在不同医学图像分割任务中的适应性，研究者们不断探索和创新，提出了众多U-Net的变体模型。这些变体模型或通过添加新的功能模块，或通过优化网络结构，以期在医学图像分割领域取得更佳的表现。例如，香草U-Net通过引入通道/空间注意力机制或自注意力机制，以捕捉医学图像中的关键信息，从而显著提升了其在多种分割任务中的性能。此外，研究者们还通过优化香草U-Net的编码器-解码器结构或调整跳跃连接，生成更为精细和丰富的特征表示。
{To further enhance the adaptability of U-Net for different medical image segmentation tasks, researchers have continuously explored and innovated, proposing numerous variant models of U-Net. These variant models aim to achieve better performance in medical image segmentation by adding new functional modules or optimizing the network structure. For instance, Vanilla U-Net introduces channel/spatial attention mechanisms or self-attention mechanisms to capture crucial information in medical images, significantly improving its performance in various segmentation tasks. Additionally, researchers have optimized the encoder-decoder structure of Vanilla U-Net or adjusted the skip connections to generate more refined and abundant feature representations.}

%然而，这些方法的引入也带来了新的问题。新增的参数和功能模块虽然提升了模型的性能，但同时也增加了模型的复杂度，提高了过拟合的风险。更为重要的是，这些方法往往忽视了粗粒度和细粒度语义信息的互补性和融合性。大多数现有的语义分割方法都假设整个分割过程可以通过单一的前馈过程来完成，从而提取到的特征表示具有同质性，难以在提取细粒度特征表示方面发挥出色的性能。因此，对于计算资源有限的医疗环境而言，如何在确保模型简洁性的基础上，充分融合与利用不同尺度的语义信息，并同时保持较小的参数量，将是极为有益的。这样的设计不仅能提升模型的泛化能力和鲁棒性，还能确保在实际应用中的高效性和实用性。
{However, the introduction of these methods has also brought new challenges. Although the addition of new parameters and functional modules enhances model performance, it also increases model complexity and the risk of overfitting. More importantly, these methods often overlook the complementarity and fusion of coarse-grained and fine-grained semantic information. Most existing semantic segmentation methods assume that the entire segmentation process can be completed through a single feedforward process, resulting in homogeneous feature representations that struggle to excel in extracting fine-grained feature representations. Therefore, for medical environments with limited computational resources, it is highly beneficial to ensure model simplicity while fully integrating and utilizing semantic information at different scales while maintaining a small number of parameters. Such a design can not only enhance the generalization and robustness of the model but also ensure its efficiency and practicality in real-world applications.}

%贡献
%在本文中，我们对医学图像分割采取了另一种立场。我们在U-Net结构中将特征学习过程划分为两个不同深度的“步骤”，以实现不同大小感受野的结合，并使网络能够学习不同粒度的语义信息。整个分割过程通过两个前馈过程完成（被称为“阶段”）。在每个阶段结束时，从该阶段获得的特征被传递到下一阶段进行融合。这种转移操作允许模型根据其在前一训练阶段学到的知识挖掘更细粒度的信息，使其能够细化粗略的分割输出。与之前的工作相比，我们提出的方法没有向U-Net添加任何额外的参数或功能模块。相反，通过内部渐进式学习来重用构建块以细化特征，并使用外部渐进式学习策略来优化每个阶段的参数并融合粗粒度和细粒度信息。从所有阶段提取的特征仅在最后一步连接，以确保充分探索特征之间的互补关系。我们的主要贡献可以总结为：
%我们提出了一种针对医学图像分割任务设计的渐进式学习网络（PL-Net）。该网络通过独特的设计，深入探索了特征互补和融合在医学图像分割中的潜力。同时，PL-Net没有额外增加功能模块是其网络架构更加简洁。
%我们引入了内部渐进式学习（IPL）和外部渐进式学习（EPL）策略。IPL策略有效捕获了不同大小的感受野，从而学习并整合多粒度的语义信息。EPL策略使得模型能够基于前阶段的知识，挖掘更精细的信息，进而优化分割结果。
%我们将提出的方法应用在皮肤病灶分割和细胞核等分割任务上，实验结果表明，PL-Net优于U-Net、BiO-Net等其他最先进的方法。此外，尽管较小版本的PL-Net†参数量少，但其分割性能依然优越。

{In this paper, we propose a new medical image segmentation method called progressive learning networks (PL-Net). PL-Net divide the feature learning process within the U-Net architecture into two distinct depth "steps" to achieve the combination of different receptive field sizes, enabling the network to learn semantic information at varying granularities. The entire segmentation process is performed through two feedforward processes (referred to as "stages"). At the end of each stage, the features obtained from that stage are transferred to the next stage for fusion. This transfer operation allows the model to leverage the knowledge learned in the previous training stage to extract finer-grained information, thereby refining the coarse segmentation output. Unlike previous works, our proposed method does not add any additional parameters or functional modules to the U-Net. Instead, our method fully explores the complementary relationships between features through a progressive learning strategy. The main contributions can be summarized as follows:}

{1) We propose a progressive learning network (PL-Net) designed specifically for medical image segmentation tasks. Through its unique design, this network deeply explores the potential of feature complementarity and fusion in medical image segmentation. By adjusting the scale of output channels, we designed both a standard PL-Net (15.03 M) and a smaller version, PL-Net† (Ocs=0.5, 3.77 M), to accommodate medical scenarios with different computational resources.}

{2) We introduce internal progressive learning (IPL) and external progressive learning (EPL) strategies. The IPL strategy effectively captures different receptive field sizes, thereby learning and integrating multi-granularity semantic information. The EPL strategy allows the model to extract finer information based on the knowledge from the previous stage, thus optimizing the segmentation results.}

{3) We applied the proposed method to tasks such as skin lesion segmentation and cell nucleus segmentation. Experimental results indicate that PL-Net outperforms other state-of-the-art methods such as U-NeXt and BiO-Net. Moreover, despite its smaller parameter size, the smaller version of PL-Net† still demonstrates superior segmentation performance.}

{\section{Releat Work}}

{Currently, most semantic segmentation methods assume that the entire segmentation process can be executed through a single feedforward pass of the input image, which often overlooks global information. To address this, researchers have added new functional modules or optimized the U-Net structure to achieve performance improvements. These methods can be classified into: 1) U-Net variants focused on functional optimization; 2) U-Net variants focused on structural optimization.}

\textbf{U-Net variants focused on functional optimization.} Due to the large number of irrelevant features in medical images, it is crucial to focus on target features and suppress irrelevant features during the segmentation process. Recent works have extended U-Net by adding different novel functional modules, demonstrating its potential in various visual analysis tasks. Squeeze-and-Excitation (SE) \citep{ref9} has facilitated the development of U-Net by automatically learning the importance of each feature channel through an attention mechanism. Additionally, ScSE \citep{ref10} and FCANet \citep{ref4} integrate concurrent spatial and channel attention modules into U-Net to improve segmentation performance. Oktay et al. \citep{ref11} proposed an attention gate for medical imaging to focus on target structures of different shapes and sizes and suppress irrelevant areas of the input image. In addition to plug-and-play attention modules, researchers have designed specific functional modules for different medical image segmentation tasks. For example, Zhou et al. \citep{ref12} proposed a contour-aware information aggregation network with a multi-level information aggregation module between two task-specific decoders. The SAUNet \citep{ref13} uses both a secondary shape stream and a regular texture stream in parallel to capture rich shape-related information, enabling multi-level interpretation of the external network and reducing the need for additional computations. The CE-Net \citep{ref14} uses a dense atrous convolution (DAC) block to extract a rich feature representation and residual multi-kernel pooling (RMP) operation to further encode the multi-scale context features extracted from the DAC block without additional learning weights.

The emergence of the Vision Transformer (ViT) \citep{ref15} has had a significant impact on the progress of medical image analysis. Compared to CNN methods, ViT has less inductive bias. The U-Transformer \citep{ref16} takes inspiration from ViT and incorporates multi-head self-attention modules into U-Net, which helps to obtain global contextual information. The UNeXt \citep{ref17} is the first fast medical image segmentation network that uses both convolution and MLP. It reduces the number of parameters and computational complexity by using tokenized MLP. In contrast to the aforementioned U-Net variants, our work explores the effectiveness of progressive learning techniques in capturing both coarse-grained and fine-grained semantic information. The PL-Net enhances the performance of different stage U-Nets by reusing learned features.

\textbf{U-Net variants focused on structural optimization.} {Unlike U-Net variants focused on functional optimization, optimizing its structure allows it to extract feature information at different levels, which is feasible and effective for many computer vision problems.} One of the simplest and most effective ways to optimize the encoder-decoder structure is to replace the basic building blocks with more advanced ones, such as \citep{ref18,ref19,ref20}, which benefit from residual or dense connections in deeper network structures. In addition to replacing the building blocks, performance can also be improved for different tasks by increasing the number of U-shaped network structures, as demonstrated in \citep{ref20,ref21}. One of the most famous networks in this category is nnU-Net \citep{ref21}, which proposes three networks based on the original U-Net structure: 2D U-Net, 3D U-Net, and U-Net cascade. The first stage performs coarse segmentation of downsampled low-resolution images, and the second stage combines the results of the first stage for fine-tuning. ResGANet \citep{ref23} achieved segmentation performance improvement by replacing the encoder in U-Net with a lightweight and efficient backbone. TransUNet \citep{ref24} and FATNet \citep{ref25} replaced the encoder structure of U-Net with CNN and Transformer branches in a parallel or serial manner.

Skip connections are considered a key component of U-Net's success. U-Net++ \citep{ref26} has redesigned skip connections through a series of nested and dense connections, reducing the semantic gap between the subnet feature map encoders and decoders. R2U-Net \citep{ref27} effectively increases the network depth by utilizing residual networks and RCNN and obtaining more expressive features through feature summation with different time steps. Xiang et al. designed BiO-Net \citep{ref28} with backward skip connections based on R2UNet, which can reuse the features of each decoding level to achieve more intermediate information aggregation. The emergence of BiO-Net allows building blocks to be reused by U-Net in a circular manner without introducing any additional parameters.

\begin{figure}
    \centering
    \includegraphics[width=0.9\textwidth]{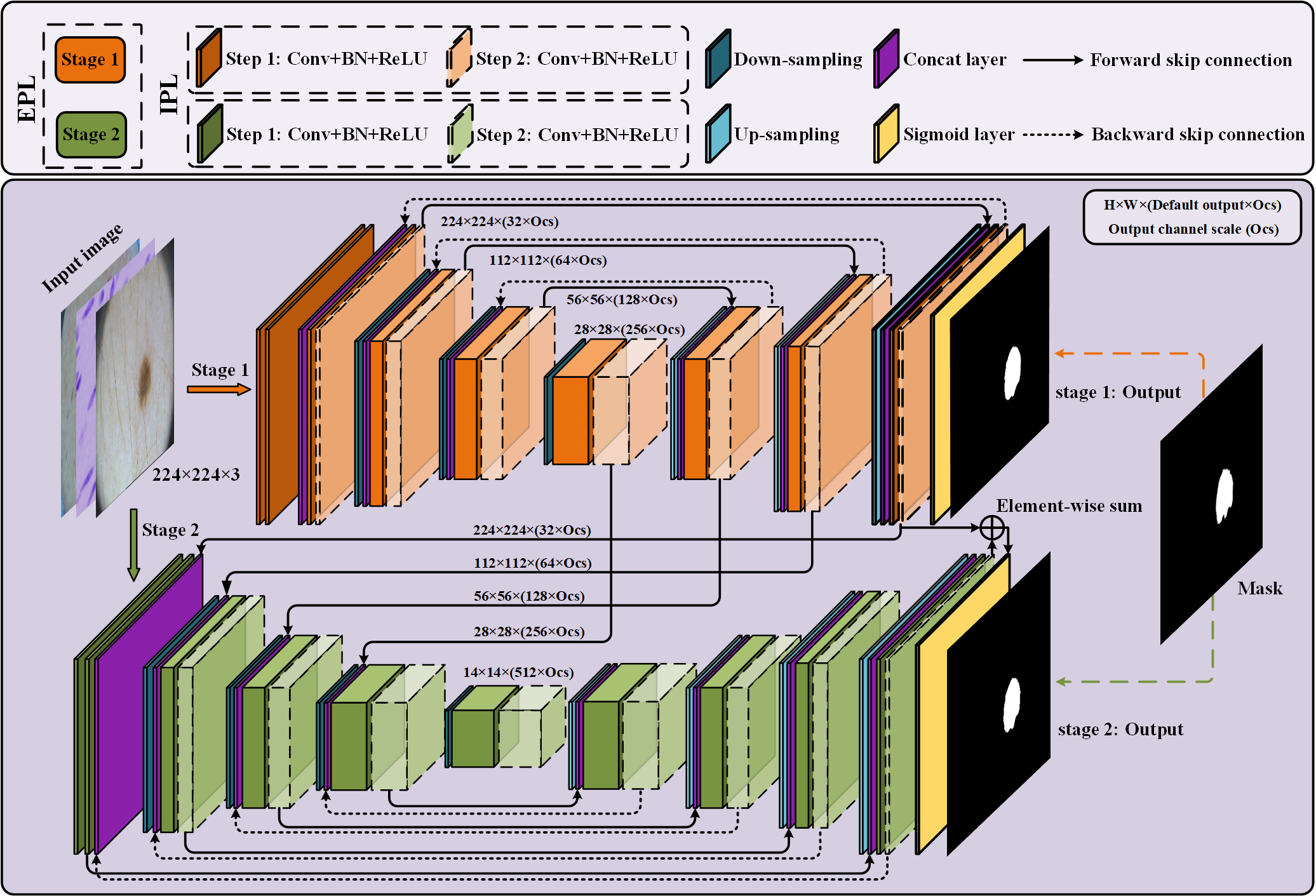}
    \caption{Overview of the progressive learning network (PL-Net). PL-Net consists of two parts: internal progressive learning (IPL) and external progressive learning (EPL).}
    \label{fig1}
\end{figure}

\section{Progressive Learning Network}

We now describe PL-Net, a progressive learning framework for medical image segmentation. As is shown in Fig. \ref{fig1}, PL-Net is a multi-level U-Net network architecture that does not rely on additional functional modules but has paired bidirectional connections. The core of our framework is to enhance the feature representation required for image segmentation through two progressive learning approaches (internal and external) and to fuse coarse-grained as well as fine-grained semantic information.

%Two U-Nets of different depths form different learning "stages". At each stage, as the "steps" increase, the shallow network is expanded into a deeper one, from which stable multi-granularity information is learned. In brief, the number of parameters is not increased through internal progressive learning, but it can learn feature maps with different sizes of receptive fields; external progressive learning is defined as the coarse segmentation stage (stage 1), and the refined segmentation stage (stage 2); %internal progressive learning is defined as the segmentation step (step 1, step 2 and so on).The input image will be checked on multiple scales to achieve the fusion of coarse-grained and fine-grained information.

{Two U-Nets with different depths form the "stages" of external progressive learning. In each stage, as the "step" of internal progressive learning increases, the shallow network is expanded to a deeper network, learning stable multi-granularity information from it. In brief, the number of parameters is not increased through internal progressive learning, but it can learn feature maps with different sizes of receptive fields. External progressive learning is defined as the coarse segmentation stage (Stage 1) and the fine segmentation stage (Stage 2). The input image will be examined at multiple scales to achieve the fusion of coarse-grained and fine-grained information.}

\subsection{Internal Progressive Learning}

Bidirectional skip connections are used in internal progressive learning to reuse building blocks. In order to enable the network at each stage to learn distinctive feature representations, we use two "steps" to gradually mine the features from shallow to deep.

Forward Skip Connections (FSC) are used to assist up-sampling learning, restore the contour of the segmentation target, and retain the low-level visual features of encoding. The feature \( f^{FSC}_{s} \) after FSC can be expressed as:

\begin{equation}
	f^{FSC}_{s}=\left [ Conv_s \left ( x \right ) ,\hat{x}  \right ] \label{XX}
\end{equation}

{Backward Skip Connections (BSC) are used for flexible aggregation of low-level visual features and high-level semantic features.} In order to realize the complementation and fusion of semantic information at different stages, multi-granularity information of different "steps" and "stages" is combined in feature \( f^{BSC} \):

\begin{equation}
f^{BSC}=\left\{\begin{matrix} 
	[x,x]    &s=1,stage1\\
	[Conv_{s}(f^{FSC}_{s}),x] &s=1,stage2 \\
	[Conv_{s}(\hat{x}),x] &s=2,stage1,2
\end{matrix}\right. 
\end{equation}

{Among them, $[\cdot]$ is the concatenation layer, $ Conv_{s} $ means that the convolution operation of the \emph{s}-th "steps" ($ s\in \left \{ {1,2} \right \} $) is applied to the input feature map, $ x $ and $ \hat{x} $ are feature maps of the same size in the down-sampling and up-sampling path respectively.}

It is worth noting that the reasoning path of internal progressive learning can be extended to multiple recursions to obtain instant performance gains. More importantly, a larger receptive field will be got in each output of this learning strategy than the previous "steps". We use $ K_{i} $() to represent the \emph{i}-th complete encoding-decoding process, and $ x_{out}^{i} $  is used to represent the output. Therefore, $ x_{out}^{i} $ can be written as:
\begin{equation}
x_{out}^{i}=\left\{\begin{matrix} 
	x_{in}    &i=0&\\ 
	K_i(x_{in}) &i=1&\\ 
	K_i(x_{out}^{i-1})&i\ge  2&
\end{matrix}\right.
\end{equation}

In this study, we define $ i=2 $, and through such a setting the parameters equivalent to BiO-Net can be maintained. In future research, the setting of $ i>2 $ can be used to further improve the segmentation accuracy, but the exploration of the optimal hyperparameter setting is beyond the scope of this paper.

\subsection{External Progressive Learning}
The external progressive learning strategy first trains the low stage (stage 1), and then gradually trains toward the high stage (stage 2). Since "stage1" is relatively shallow in depth and limited by the perceptual field and performance ability, it will focus on local information extraction, while "stage 2" incorporates the local texture information learned from "stage 1". Compared with directly training the entire network in series, in the model, it is allowed by this incremental nature to pay attention to global information as the features gradually enter a higher stage.

For each stage of training, we calculate the loss based on the Dice coefficient ($ \mathcal{L}_{Dice} $) \citep{ref29} between the ground truth ($ y_{true} $) and the predicted probability ($ y_{pred}^{n} $) distribution of different stages:
\begin{equation}
\mathcal{L}_{Dice}(y_{pred}^{n},y_{true})=1-\frac{2\times\vert y_{pred}^{n}\cap y_{true}\vert}{\vert y_{pred}^{n} \vert + \vert y_{true} \vert}
\end{equation}

Here $ \vert \cdot \vert $ is an operator through which the number of pixels is found in the qualified area. In each iteration, the input data will be used in each learning stage (where $  n\in \left \{ {1,2} \right \} $). What needs to be clear is that when the latter stage is predicted, all the parameters of the previous stage are optimized and updated, which helps each stage in the model to work together.

Since the low stage is mainly to assist the feature expression and knowledge reasoning of the high-stage network, we can delete the low-stage prediction layer (Sigmoid layer) when predicting, thereby reducing the reasoning time. In addition, the predictions at different stages are unique, but they can form complementary information among different granularities. When we combine all outputs with an equal weight, it will result in a better performance. In other words, the final output of the model is determined by all stages:
\begin{equation}
y=\frac{1}{1+e^{-\sum_{n=1}^{N-1}y^{n}}} 
\end{equation}

\begin{table*}[ht]
        \begin{center}
	\renewcommand{\tablename}{Table}
	\caption{Detailed configuration of U-Net, BiO-Net, and our PL-Net architecture. We use "[kernel, kernel, channel]" to represent the convolution configuration}
	\resizebox{1.0\textwidth}{!}{
		\begin{tabular}{l|ccc|c|ccc} \hline %\toprule[1tp]
			\multirow{2}{*}{Input} & \multicolumn{3}{c|}{Encoder}                                                                                                            & \multirow{2}{*}{Output} & \multicolumn{3}{c}{Decoder}                                                                                                           \\ \cline{2-4} \cline{6-8} 
			& U-Net          & BiO-Net        & PL-Net                                                                         &                         & U-Net           & BiO-Net         & PL-Net   \\ \hline
			$224^2$                    & {[}3, 3, 64{]}×2   & {[}3, 3, 32{]}×2  & $\left \{ \begin{tabular}[c]{@{}c@{}}{[}3, 3, 32{]}, step1\\ {[}3, 3, 32{]}, step2\\ stage1, stage2\end{tabular}\right \} $   & $7^2$                      & —               & {[}3, 3, 256{]}×2 & —                                                                                                 \\ \hline
			$112^2$                    & {[}3, 3, 128{]}×2  & {[}3, 3, 32{]}×2  & $\left \{ \begin{tabular}[c]{@{}c@{}}{[}3, 3, 64{]}, step1\\ {[}3, 3, 64{]}, step2\\ stage1, stage2\end{tabular}\right \} $   & $28^2$                     & {[}3, 3, 512{]}×2 & {[}3, 3, 128{]}×2 & $\left \{\begin{tabular}[c]{@{}c@{}}{[}3, 3, 256{]}, step1\\ {[}3, 3, 256{]}, step2\\ stage2\end{tabular}\right \} $        \\ \hline
			$56^2$                     & {[}3, 3, 256{]}×2  & {[}3, 3, 64{]}×2  & $\left \{ \begin{tabular}[c]{@{}c@{}}{[}3, 3, 128{]}, step1\\ {[}3, 3, 128{]}, step2\\ stage1, stage2\end{tabular}\right \} $ & $56^2$                     & {[}3, 3, 256{]}×2 & {[}3, 3, 64{]}×2  & $\left \{ \begin{tabular}[c]{@{}c@{}}{[}3, 3, 128{]}, step1\\ {[}3, 3, 128{]}, step2\\ stage1, stage2\end{tabular}\right \} $ \\ \hline
			$28^2$                     & {[}3, 3, 512{]}×2  & {[}3, 3, 128{]}×2 & $\left \{ \begin{tabular}[c]{@{}c@{}}{[}3, 3, 256{]}, step1\\ {[}3, 3, 256{]}, step2\\ stage1, stage2\end{tabular}\right \} $ & $112^2$                    & {[}3, 3, 128{]}×2 & {[}3, 3, 32{]}×2  & $ \left \{\begin{tabular}[c]{@{}c@{}}{[}3, 3, 64{]}, step1\\ {[}3, 3, 64{]}, step2\\ stage1, stage2\end{tabular}\right \} $   \\ \hline
			$14^2$                     & {[}3, 3, 1024{]}×2 & {[}3, 3, 256{]}×2 & $\left \{ \begin{tabular}[c]{@{}c@{}}{[}3, 3, 512{]}, step1\\ {[}3, 3, 512{]}, step2\\ stage2\end{tabular}\right \} $        & $224^2$                    & {[}3, 3, 64{]}×2  & {[}3, 3, 32{]}×2  & $ \left \{\begin{tabular}[c]{@{}c@{}}{[}3, 3, 32{]}, step1\\ {[}3, 3, 32{]}, step2\\ stage1, stage2\end{tabular}\right \} $   \\ \hline
			$7^2$                      & —                & {[}3, 3, 512{]}×2 & —                                                                                                 & $224^2$                    & \multicolumn{3}{c}{{[}1, 1, 1{]}, $ Sigmoid $}                                                                                               \\ \hline
			\multicolumn{5}{l}{Parameters}                                                                                                                                                             & 25.59 M         & 14.30 M         & 15.03 M                                                                                           \\ \hline
			\multicolumn{5}{l}{Model size}                                                                                                                                                             & 118 MB          & 57.7 MB         &60.7MB                                                                                           \\ 
			\hline%\bottomrule[1pt]                
	\end{tabular}}
	\label{Tab1}
 \end{center}
\end{table*}

\subsection{PL-Net Architecture}
Our framework has a trade-off between performance and parameters. Like U-Net, the down-sampling and up-sampling stages of PL-Net only use standard convolutional layers, batch normalization layers and ReLU layers without introducing any additional functional modules. Table.\ref{Tab1} is the detailed configuration of U-Net, BiO-Net and our PL-Net.

As shown in Table.\ref{Tab1}, BiO-Net has a maximum coding depth of 4, using BSC from the deepest coding level, and inputting the decoded features in each iteration as a whole into the last-stage block. BSC is also used in PL-Net. Unlike the previous methods, the convolutional layer is allowed to be used in the model to mine features from coarse-grained to fine-grained ones in a progressive manner. It should be noted that a smaller version of PL-Net† can be obtained only by adjusting the Ocs, whose depth and connection method will not change.
% For Original Research articles, please note that the Material and Methods section can be placed in any of the following ways: before Results, before Discussion or after Discussion.

\section{Experiments}

\subsection{Datasets}
\textbf{ISIC 2017} \citep{ref3} is a dataset consisting of 2000 training images, 150 validation images, and 600 test images. The images in the original dataset provided by ISIC have different resolutions. To address this, we first use the gray world color consistency algorithm to normalize the colors of the images and then adjust the size of all images to 2242 pixels. All experimental results reported in this paper for ISIC 2017 are from the official test set results. 

\textbf{PH2} \citep{ref30} is a dataset containing 200 dermoscopic images, with a fixed size of 768×560 pixels. The dataset contains 80\% benign mole cases and 20\% melanoma cases, with ground truth annotated by dermatologists. Due to the small scale of the dataset, we use the preprocessing method of the ISIC 2017 dataset and the trained model to directly predict all images in the dataset to evaluate the performance of different models.

\textbf{Kaggle 2018 Data Science Bowl} (referred to as Nuclei) \citep{ref31} is a dataset provided by the Booz Allen Foundation, containing 670 cell feature maps with ground truth for each image. To prepare the dataset for training and testing, we adjust all images and corresponding ground truth to 2242 pixels and use 80\% of the images for training and the remaining 20\% for testing.

The \textbf{TN-SCUI} \citep{ref32} dataset is a collection of 3644 nodular thyroid images, each annotated by experienced physicians. The dataset was originally part of the TN-SCUI 2020 challenge and was processed to remove personal labels to protect patient privacy. In this study, we randomly divided the dataset into a training set (60\%), validation set (20\%), and test set (20\%). To ensure consistency, we uniformly adjusted the resolution of all images to 2242 pixels.

\begin{figure}
    \centering
    \includegraphics[width=0.9\textwidth]{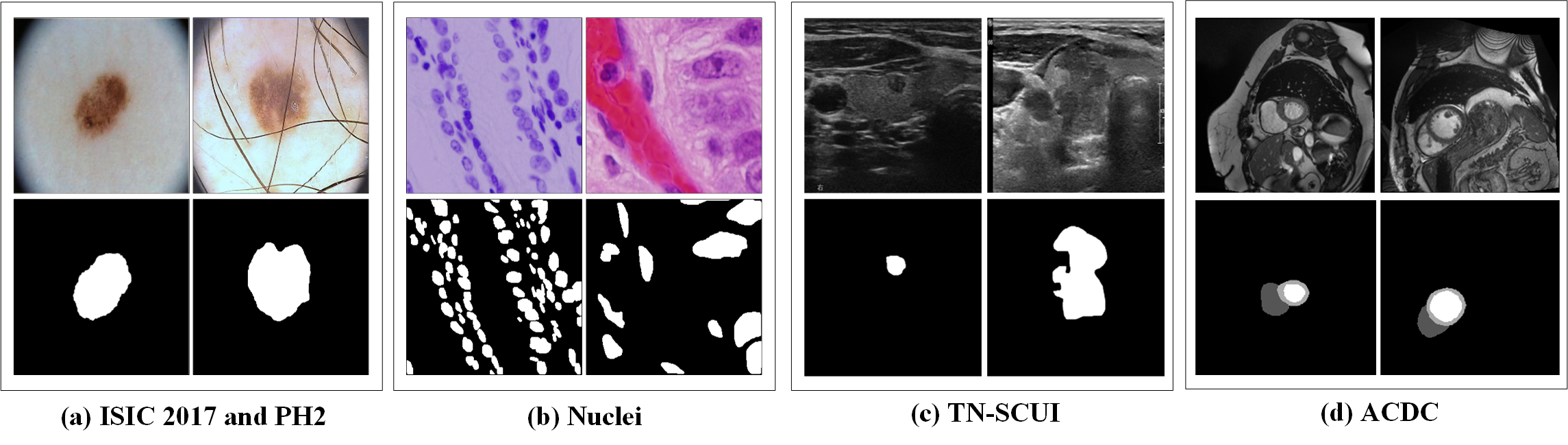}
    \caption{(a)-(d) represent samples from the five datasets.}
    \label{fig2}
\end{figure}

\textbf{ACDC} \citep{ref5} is a dataset that includes cardiac MRI images of 150 patients, from which we collected 1489 slices for 3D images. For training and testing purposes, we used 951 and 538 slices, respectively. Notably, in contrast to the four other datasets mentioned earlier, ACDC comprises three different categories: left ventricle, right ventricle, and myocardium. Hence, we employed this dataset to investigate how various models perform on multi-class segmentation. Fig. \ref{fig2} displays sample images from these datasets and their corresponding ground truth.

\subsection{Implementation Details}
We conducted all experiments on Tesla V100 GPUs using Keras and expanded the training data for all datasets by applying random rotations (±25°), random horizontal and vertical shifts (15\%), and random flips (horizontal and vertical). For all models, we trained for more than 200 epochs with a batch size of 16, a fixed learning rate of 1e-4, and an Adam optimizer with a momentum of 0.9 to minimize Dice loss. We used an early stop mechanism and stopped training when the validation loss reached a stable level with no significant change for 20 epochs. Unless explicitly specified, PL-Net had two "steps" and "stages", and BSC was established at each stage of the network. When testing, all prediction layers are deleted before the last "stage", and other configurations remain unchanged.
%There are 5 heading levels
\begin{table*}[ht]\small\centering
	\renewcommand{\tablename}{Table}
	\centering
	\caption{Ablative results. IoU (Dice), number of parameters, and model size are reported.}
 \resizebox{0.9\textwidth}{!}{
	\begin{tabular}{l|c|cccc}
	\hline%\toprule[1pt]

\multirow{2}{*}{Dataset}   & {EPL} & \multicolumn{2}{c}{Without IPL} & \multicolumn{2}{c}{With IPL}  \\ \cline{3-6} 
&                      & n=1            & n=2            & n=2           & n=3           \\ \hline
\multirow{2}{*}{ISIC 2017} & ×                    & 76.05 (84.43)  & 77.09 (85.23)  & 77.15 (85.37) & 77.07 (85.44) \\
& \checkmark                    & 76.69 (84.94)  & 77.04 (85.27)  &\textbf{77.92} (\textbf{85.94}) & 77.49 (85.56) \\ \hline
\multirow{2}{*}{Nuclei}    & ×                    & 85.54 (91.89)  & 85.13 (91.53)  & 85.93 (\textbf{92.14}) & 84.78 (91.28) \\
& \checkmark                    & 85.60 (91.84)  & 85.80 (92.00)  & \textbf{86.14} (92.13) & 85.37 (91.70) \\ \hline
\multirow{2}{*}{PH2}       & ×                    & 83.90 (90.74)  & 85.88 (91.61)  & 86.69 (92.47) & 86.48 (92.45) \\
& \checkmark                    & 84.97 (91.00)  & 86.63 (92.44)  & \textbf{87.27}  (\textbf{92.86}) & 87.03 (92.77) \\ \hline
\multirow{2}{*}{TN-SCUI}       & ×                    & 72.32 (81.38)  & 73.72 (82.61)  & 75.95 (85.36) & 76.67 (84.60) \\
& \checkmark                    & 74.20 (83.33)  & 75.63 (84.33)  & 76.66 (85.10) & \textbf{77.05} (\textbf{85.55}) \\ \hline
\multirow{2}{*}{ACDC}       & ×                    & 74.44 (81.56)  & 74.66 (82.03)  & 77.78 (83.84) & 77.49 (83.60) \\
& \checkmark                    & 75.30 (82.19)  & 76.80 (83.42)  & \textbf{78.06} (\textbf{84.36}) & 77.96 (83.91) \\ \hline
Parameters                 &—                      & \textbf{10.33M}       & 15.03 M        & 15.03 M       & 19.73 M       \\ \hline
Model size                 &—                      & \textbf{41.60MB}        & 60.70 MB       & 60.70 MB      & 79.70 MB      \\ 
\hline%\bottomrule[1pt]
\end{tabular}}
\label{Tab2}
\end{table*}

\subsection{Ablation Study}

To understand the effectiveness of IPL and EPL strategies, we conducted ablation studies. When there is no IPL strategy, features are extracted by naturally stacking benchmark blocks, and we conducted experiments on stacking 1-layer and 2-layer benchmark blocks, respectively. Adopting an IPL strategy means that the encoder-decoder must be iterated for n times in different stages, and we set n=2 and n=3. When external progressive learning is not performed, different "stages" are connected in series through PL-Net to transfer the feature information learned in each stage. Only the parameters in the last stage are optimized, and the segmentation results are output through the model. That is to say, the feature information obtained in the current "stage" of training is transferred to the next training "stage" and fused through the EPL method, allowing fine-grained information to be mined through the model based on learning in the previous training "stage".

\begin{figure}[ht]
    \centering
    \includegraphics[width=0.9\textwidth]{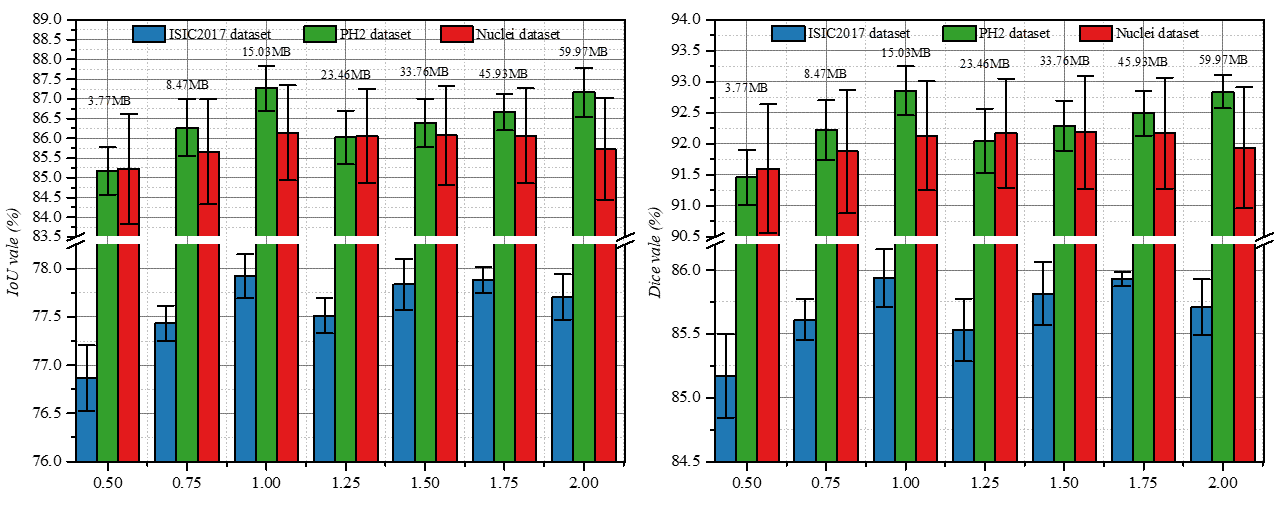}
    \caption{Study the impact of Ocs on three public datasets. Results are calculated over 5 runs and are shown with standard errors. We label the parameters of the model at the top of the bar chart.}
    \label{fig3}
\end{figure}

Table.\ref{Tab2} presents our IoU (Dice) scores without/with a progressive learning strategy on five different medical image segmentation datasets. We provide the parameters and model sizes for different scenarios to comprehensively analyze segmentation performance. In most cases, the best segmentation performance is achieved through PL-Net when both internal (n=2) and external progressive learning strategies are used simultaneously. Compared to the model with the same parameter settings without IPL, the segmentation performance is significantly improved. These results demonstrate the effectiveness of EPL and IPL.
Moreover, we observed that the progressive learning strategy has a significant impact on datasets with complex boundaries or multi-category datasets. On the TN-SCUI dataset, for instance, the IoU improvement is as high as 2.94\% with the same parameter setting (n=2). To balance factors such as performance and parameters, we used the setting of n=2 in the following experiments. However, we believe that the setting of n=3 may be more effective as the size of the dataset increases.

\begin{table*}[ht]\small\centering
	\renewcommand{\tablename}{Table}
	\caption{Performance comparison with SOTA methods on ISIC 2017 and PH2 datasets. {\color{red} Red}, {\color{green} Green}, and {\color{blue} Blue} indicate the best, second best and third best performance.}
	\resizebox{\textwidth}{!}{
		\centering\begin{tabular}{l|ccccc|ccccccc}
			\hline%\toprule[1pt]
			\multirow{2}{*}{Network}    & \multicolumn{5}{c|}{ISIC 2017 Dataset} & \multicolumn{5}{c}{PH2 Dataset}       & {\#Params} & {Model size} \\ \cline{2-11}
			& Acc    & IoU   & Dice  & Sens  & Spec  & Acc   & IoU   & Dice  & Sens  & Spec  &                           &                             \\ \hline
			FrCN {\citep{ref33}}               & {\color{red} 0.940}  & 0.771 & {\color{red} 0.871} & 0.854 & 0.967 & 0.951 & 0.848 & 0.918 & 0.937 & 0.957 & —                         & —                           \\
			FocusNet {\citep{ref7}}            & 0.921  & 0.756 & 0.832 & 0.767 & {\color{red} 0.990} & —     & —     & —     & —     & —     & —                         & —                           \\
			SegNet {\citep{ref34}}             & 0.918  & 0.696 & 0.821 & 0.801 & 0.954 & 0.934 & 0.808 & 0.894 & 0.865 & {\color{green} 0.966} & 28.09M                    & 112 MB                      \\
			DSNet {\citep{ref14}}              & —      & {\color{blue} 0.775} & —     & {\color{red} 0.875} & 0.967 & —     & {\color{blue} 0.870} & —     & 0.929 & {\color{red} 0.969} & {\color{green} 10.00 M}                   & —                           \\
			DAGAN {\citep{ref17}}              & 0.935  & 0.771 & {\color{green} 0.859} & 0.835 & {\color{green} 0.976} & —     & —     & —     & —     & —     & —                         & —             \\            
                FATNet  {\citep{ref25}}              & 0.933  & 0.765 & 0.850 & 0.839 & 0.973 & —     & —     & —     & —     & —     & 27.43 M                         & 109 MB                  \\     
                ResGANet  {\citep{ref23}}              & 0.936  & 0.764 & {\color{green} 0.862} & 0.842 & 0.950 & —     & —     & —     & —     & —     & 39.21M                         & —                           \\ \hline
			U-Net {\citep{ref2}}               & 0.926  & 0.736 & 0.825 & 0.828 & 0.964 & 0.943 & 0.851 & 0.915 & 0.946 & 0.957 & 29.59 M                   & 118 MB                      \\
			U-Net++ {\citep{ref3}}             & 0.929  & 0.753 & 0.840 & 0.848 & 0.965 & 0.948 & 0.853 & 0.917 &  {\color{red} 0.973}  & 0.937 & 34.48 M                   & 138 MB                      \\
			Double U-Net {\citep{ref21}}       & 0.936  & 0.765 & 0.847 & 0.830 & 0.970 & 0.942 & 0.860 & 0.915 & 0.934 & 0.953 & 27.94 M                   & 112 MB                      \\
			FCANet {\citep{ref9}}              & 0.935  & {\color{green} 0.776} & 0.856 & {\color{blue} 0.869} & 0.962 & {\color{blue} 0.952} & 0.868 & 0.926 & {\color{green} 0.968} & 0.926 & 59.97 M                   & 241 MB                      \\
			Att U-Net {\citep{ref4}}           & {\color{green} 0.939}  & 0.757 & 0.840 & 0.859 & 0.957 & 0.951 & 0.868 & 0.926 & 0.953 & 0.955 & 30.42 M                   & 121 MB                      \\
			R2U-Net {\citep{ref23}}            & {\color{blue} 0.938}  & {\color{green} 0.776} & {\color{blue} 0.858} & 0.859 & 0.969 & 0.952 & {\color{green} 0.871} & {\color{blue} 0.927} & 0.954 & 0.960 & 91.61 M                   & 366 MB                      \\
			Att R2U-Net {\citep{ref23}}        & {\color{green} 0.939}  & {\color{blue} 0.775} & 0.857 & 0.857 & 0.961 & {\color{green} 0.954} & {\color{red} 0.873} & {\color{green} 0.928} & 0.949 & 0.962 & 92.11 M                   & 368 MB                      \\
			BiO-Net (t=3) {\citep{ref26}}      & 0.937  & 0.772 & 0.852 & 0.845 & 0.973 & 0.944 & 0.851 & 0.915 & 0.963 & 0.931 & {\color{blue} 14.30 M}                   & {\color{green} 57.7 MB}                     \\
			BiO-Net (t=3, INT) {\citep{ref26}} & 0.934  & 0.754 & 0.840 & 0.821 & {\color{green}0.976} & 0.945 & 0.851 & 0.909 & {\color{blue}0.968} & 0.944 & {\color{green} 3.99 M}                   & {\color{red} 15.2 MB}      \\               
                UNeXt-L {\citep{ref17}} & 0.935  & 0.773 & 0.856 & 0.864 & 0.966 & 0.949 & 0.859 & 0.919 & 0.941 & 0.955 & {\color{blue} 14.30 M}                   & {\color{green} 57.7 MB}                     \\ \hline
			PL-Net† (Our)               & 0.932  & 0.769 & 0.852 & {\color{green} 0.871} & 0.953 & 0.945 & 0.852 & 0.915 & 0.955 & 0.957 & {\color{red} 3.77 M}                    & {\color{red} 15.6 MB}                     \\
			PL-Net (Our)                & {\color{red} 0.940}  & {\color{red} 0.779} & {\color{green} 0.859} & 0.848 & {\color{blue} 0.975} & {\color{red} 0.957} & {\color{red} 0.873} & {\color{red} 0.929} & {\color{blue} 0.965} & {\color{green} 0.966} & 15.03 M                   & {\color{blue} 60.7 MB}                      \\ \hline%\bottomrule[1pt]
	\end{tabular}}
	\label{Tab3}
\end{table*}

In addition to the above ablation studies, we also investigated the impact of the output channel scale (Ocs) on the segmentation performance of different datasets. Fig. \ref{fig3} shows the experimental results on three datasets, where we set $Ocs \in [0.5, 2.0]$ and take values at an interval of 0.25. Note that Ocs=0.5 represents a smaller version of PL-Net†.
We found that when Ocs$=$1.0, the best segmentation result can be obtained, and the parameter amount (15.03MB) is well-balanced. When Ocs $>$ 1.0, the segmentation performance improves as the number of channels increases, but it does not exceed that of the standard PL-Net. We attribute this to the limitation of the data size and the complexity of the segmentation content. While PL-Net† has slightly lower segmentation performance than other networks, it has very few parameters. Thus, it is recommended for use on small datasets. Additionally, it can be configured to run on servers or mobile devices with lower hardware requirements.

\begin{table*}[ht]\tiny\centering
	\renewcommand{\tablename}{Table}
	\caption{Performance comparison with SOTA methods on Nuclei dataset. {\color{red} Red}, {\color{green} Green}, and {\color{blue} Blue} indicate the best, second-best, and third-best performance. For the original implementation methods, we report mean ± standard deviation.}
 \resizebox{\textwidth}{!}{
		  \begin{tabular}{l|cccccccccccc}
    			\hline%\toprule[1pt]
    			\multirow{2}{*}{Network}    & \multicolumn{4}{c}{Nuclei Dataset}        &\multirow{2}{*}{\#Params} & \multirow{2}{*}{Model size} \\ \cline{2-5}
    			& Acc    & IoU   & Dice  & Sens   &       &                             \\ \hline
    			PraNet {\citep{ref38}}               & 95.59  & 71.08 & 81.03 & 80.62  & —                         & —                           \\
    			Channel-UNet {\citep{ref39}}            & 96.27  & 79.75 & 87.55 & 90.70 & —                         & —                           \\
    			ResUNet {\citep{ref18}}             & 97.05  & 82.44 & 89.91 & 90.00  & —                         & —               \\
    			Double U-Net {\citep{ref22}}              & —      & 84.07 & 91.33     & 64.07 & 27.94 M & 112 MB                          \\
    			TransAttUnet D {\citep{ref40}}              & 97.37  & 84.62 & 91.34 & 91.86 & —                         & —             \\            
                TransAttUnet R  {\citep{ref40}}              & 97.46  & 84.98 & 91.62 & 91.85  & —     & —            \\     
                TransUNet  {\citep{ref24}}              & 97.84  & 85.21 & 91.69 & 91.62 & 100.4 M & 401 MB                     \\
                FATNet  {\citep{ref25}}              & {\color{red} 98.11}  & 85.24 & 91.69 & 91.73 & 27.43 M & 109 MB                     \\ \hline
    			U-Net {\citep{ref2}}               & 97.84±0.24 & 85.68±1.40 & 91.90±1.00 & {\color{red} 92.61±0.52} & 29.59 M & 118 MB                    \\
    			U-Net++ {\citep{ref26}}             & 97.87±0.22  & {\color{green} 85.91±1.35} & {\color{blue} 92.06±1.00} & {\color{blue} 92.48±1.08} & 34.48 M & 138 MB                     \\
    			FCANet {\citep{ref4}}       & 97.68±0.31  & 84.87±1.39 & 91.33±1.09 & 91.70±1.50 & 59.97 M & 241 MB             \\
    			Att U-Net {\citep{ref11}}              & 97.84±0.18  & 85.46±1.20 & 91.77±0.88 & 91.93±0.66 & 30.42 M & 121 MB                     \\
    			R2U-Net {\citep{ref27}}            & {\color{blue} 97.93±0.18}  & 85.68±1.26 & 91.89±0.92 & 92.28±1.20 & 91.61 M & 366 MB                      \\
    			Att R2U-Net {\citep{ref27}}        & 97.76±0.34  & {\color{blue} 85.86±1.04} &  {\color{red} 92.15±0.92} & {\color{green} 92.51±1.46} & 92.11 M & 368 MB                      \\
    			BiO-Net (t=3) {\citep{ref28}}      & 97.81±0.22  & 85.09±1.42 & 91.53±1.04 & 91.99±0.72  & {\color{blue} 14.30 M}                   & {\color{blue} 57.7 MB}                     \\
    			BiO-Net (t=3, INT) {\citep{ref28}} & 97.84±0.20  & 85.31±1.27 & 91.68±0.93 & 91.94±0.76 & {\color{blue}14.30 M} & {\color{blue} 57.7 MB}      \\               
                UNeXt-L {\citep{ref17}} & 97.43±0.15  & 81.26±1.46 & 88.75±1.31 & 88.71±1.65 & {\color{green} 3.99 M}                   & {\color{green} 15.2 MB}                     \\ \hline
    			PL-Net† (Our)               & 97.79±0.22  & 85.23±1.39 & 91.60±1.03 & 91.79±0.59  & {\color{red} 3.77 M}                    & {\color{red} 15.6 MB}                     \\
    			PL-Net (Our)                & {\color{green} 97.96±0.16} & {\color{red} 86.14±1.20}  & {\color{green} 92.13±0.88} & 92.12±1.11  & 15.03 M & 60.7 MB                    \\ \hline%\bottomrule[1pt]
        	\end{tabular}}
	\label{Tab4}
\end{table*}

\subsection{Comparison with State-of-the-arts}

\subsubsection{Quantitative Comparison}

For the \textbf{ISIC 2017} and \textbf{PH2} datasets, we compared our PL-Net to the baseline U-Net and other state-of-the-art methods \citep{ref2,ref4,ref11,ref17,ref19,ref21,ref22,ref23,ref25,ref26,ref27,ref28,ref33,ref34,ref35,ref36,ref37}. The functional optimization-oriented variants of U-Net include \citep{ref4,ref11,ref17,ref34,ref27,ref33,ref35} while the structural optimization-oriented variants of U-Net include \citep{ref19,ref21,ref22,ref23,ref25,ref26,ref27,ref28,ref36,ref37}. To ensure fairness, we either used the experimental results provided by the authors on the same test set or ran their models published in the same environment.

Table.\ref{Tab3} presents the accuracy (Acc), intersection over union (IoU), Dice coefficient (Dice), sensitivity (Sens), and specificity (Spec) scores of different segmentation methods on the ISIC2017 and PH2 datasets. Our PL-Net outperforms other methods in terms of both IoU and Dice metrics on the ISIC2017 dataset. Specifically, the IoU and Dice scores of PL-Net are 0.6\% and 0.3\% higher than those of BiO-Net (t = 3, INT), respectively. The smaller sized PL-Net† (3.77 M) achieves the same Dice score as BiO-Net (t = 3) (14.30 M). Although nnU-Net \citep{ref16} achieves the best sensitivity on the ISIC2017 test set, its model size is 3.76 times larger than that of the standard PL-Net. The PH2 dataset also involves the dermoscopic image segmentation task. While the number of parameters in UNeXt-L \citep{ref15} is similar to that of our smaller version of PL-Net†, UNeXt-L completes the entire segmentation process through a single feed-forward pass of the input image, resulting in low parameter utilization and insufficient learning. When compared with other state-of-the-art methods, PL-Net demonstrates superior performance on the PH2 dataset. Furthermore, PL-Net† hasmuch fewer parameters than other methods, yet it still achieves competitive segmentation performance.

\begin{table*}[ht]\small\centering
	\renewcommand{\tablename}{Table}
	\caption{Performance comparison with SOTA methods on TN-SCUI datasets. {\color{red} Red}, {\color{green} Green}, and {\color{blue} Blue} indicate the best, second-best, and third-best performance.}
	\setlength{\tabcolsep}{0.6mm}{
  % \resizebox{0.7\textwidth}{!}{
		  \begin{tabular}{l|cc|cc}
    			\hline%\toprule[1pt]
    			\multirow{2}{*}{Network}   & \multicolumn{2}{c|}{TN-SCUI Datasets}         & \multirow{2}{*}{\#Params} & \multirow{2}{*}{Model size} \\ \cline{2-3}
    			& IoU   & Dice     &    &                             \\ \hline
    			
    			U-Net {\citep{ref2}}               & 0.718 & 0.806  & 29.59 M & 118 MB                    \\
    			SegNet {\citep{ref35}}             & 0.726  & 0.819  & 17.94 M & 71.8 MB                    \\
    			FATNet {\citep{ref25}}        & {\color{green}0.751}  & {\color{green}0.842}  & 27.43 M & 109 MB          \\
    			Swin-UNet {\citep{ref37}}     & 0.744  & 0.835       & 25.86 M & 105 MB                    \\
    			TransUNet  {\citep{ref24}}            & {\color{blue} 0.746}  & 0.837  & 88.87 M & 401 MB                     \\
    			EANet {\citep{ref41}}        & {\color{green} 0.751}  & {\color{blue} 0.839}  & 47.07 M & 118 MB                      \\
                UNeXt-L {\citep{ref17}} & 0.693  & 0.794  & {\color{green} 3.99 M}                   & {\color{red} 15.2 MB}                     \\ \hline
    			PL-Net† (Our)               & 0.742  & 0.830    & {\color{red} 3.77 M}                    & {\color{green} 15.6 MB}                     \\
    			PL-Net (Our)                & {\color{red} 0.767} & {\color{red} 0.851}   & {\color{blue}15.03 M} & {\color{blue} 60.7 MB}                   \\ \hline%\bottomrule[1pt]
        	\end{tabular}}
	\label{Tab5}
\end{table*}

\begin{table*}[ht]\tiny\centering
	\renewcommand{\tablename}{Table}
	\caption{Performance comparison with SOTA methods on ACDC datasets. {\color{red} Red}, {\color{green} Green}, and {\color{blue} Blue} indicate the best, second-best, and third-best performance.}
	\setlength{\tabcolsep}{0.9mm}{
		  \begin{tabular}{l|cccccccc|cc}
    			\hline%\toprule[1pt]
    		   \multirow{2}{*}{Network} & \multicolumn{4}{c}{ACDC Datasets}        & \multirow{2}{*}{\#Params} & \multirow{2}{*}{Model size} \\ \cline{2-5}
    			  & RV  & Myo & LV &Average   &    &                             \\ \hline
    			
    			U-Net {\citep{ref2}}               & 0.743(0.792) & 0.717(0.812) & 0.861(0.897) & 0.774(0.834) & 29.59 M & 118 MB                    \\
    			SegNet {\citep{ref35}}             &0.738(0.790) &0.720(0.817) &{\color{blue}0.864}(0.902) &0.774(0.836) & 17.94 M & 71.8 MB                    \\
    			FATNet {\citep{ref25}}        & 0.743(0.799)  & 0.702(0.805) & 0.859(0.899) & 0.768(0.834) & 27.43 M & 109 MB          \\
    			Swin-UNet {\citep{ref37}}         & {\color{green}0.754(0.805)}  & {\color{blue}0.722(0.820)} & {\color{green}0.865(0.903)} & {\color{green}0.780(0.843)} & 25.86 M & 105 MB                    \\
    			TransUNet  {\citep{ref24}}             & {\color{blue}0.750(0.800)}  & 0.715(0.812)& {\color{red}0.872}{\color{green}(0.905)} & {\color{blue}0.779(0.839)} & 88.87 M & 401 MB                     \\
    			EANet {\citep{ref41}}         &0.742(0.791) &  {\color{green} 0.732(0.825)} & {\color{blue} 0.864}(0.902) & {\color{blue} 0.779(0.839)} & 47.07 M & 118 MB                      \\
                UNeXt-L {\citep{ref17}}  & 0.719(0.779) & 0.675(0.810) &0.840(0.882) & 0.745(0.824) & {\color{green} 3.99 M}                   & {\color{red} 15.2 MB}                     \\ \hline
    			PL-Net† (Our)                & 0.723(0.778) & 0.692(0.796) &0.845(0.887) &0.753(0.820)   & {\color{red} 3.77 M}                    & {\color{green} 15.6 MB}                     \\
    			PL-Net (Our)                 & {\color{red} 0.761(0.807)} & {\color{red} 0.738(0.828)}  & {\color{red} 0.872(0.907)} & {\color{red} 0.790(0.847)}  & {\color{blue}15.03 M} & {\color{blue} 60.7 MB}                   \\ \hline%\bottomrule[1pt]
        	\end{tabular}}
	\label{Tab6}
\end{table*}

\begin{figure}
    \centering
    \includegraphics[width=0.95\textwidth]{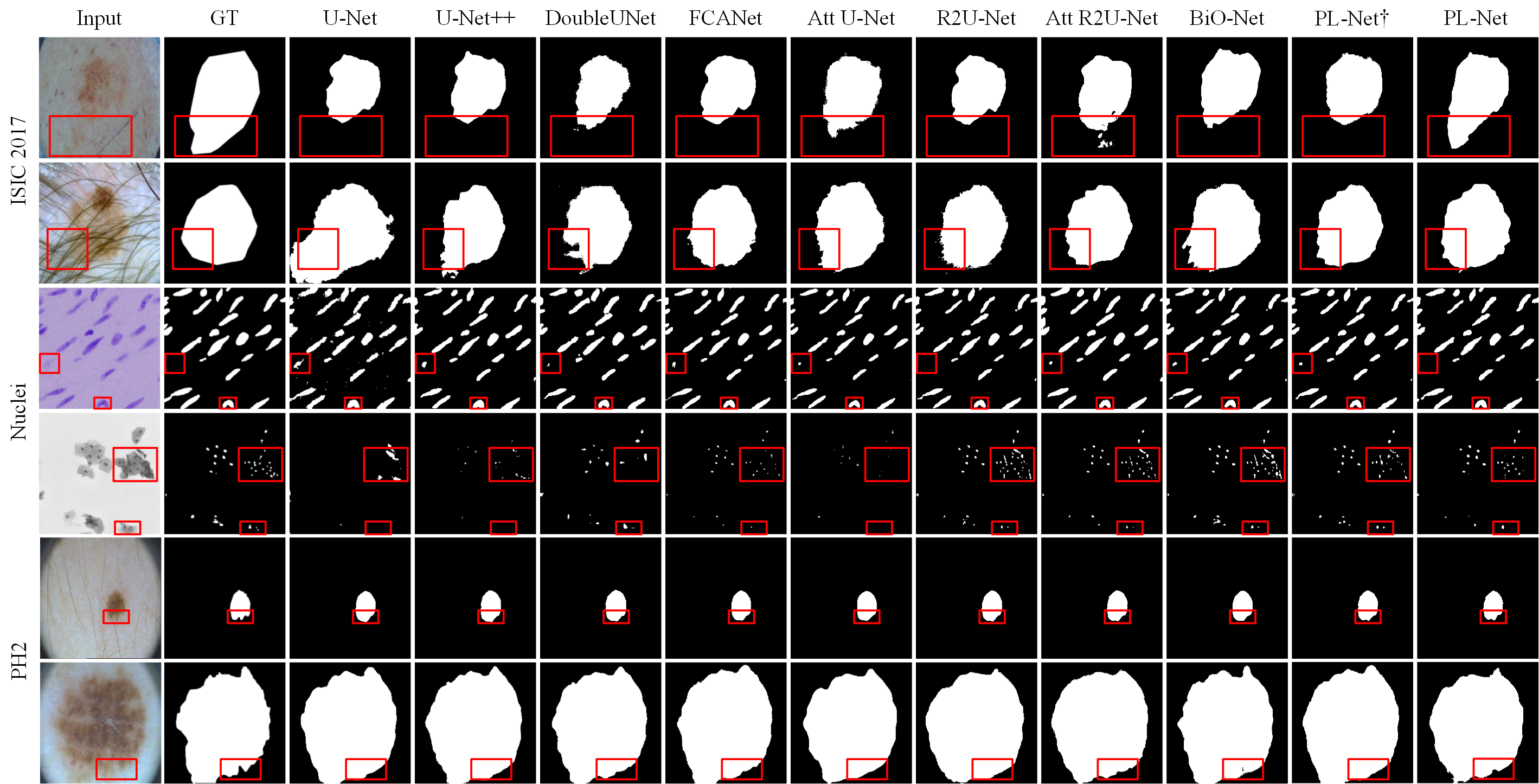}
    \caption{Qualitative segmentation results of ISIC 2017, Nuclei, and PH2 datasets using different methods.}
    \label{fig4}
\end{figure}

\textbf{Nuclei dataset.} The datasets used for nucleus segmentation have non-uniform feature distributions, and the shapes of positive and negative samples vary greatly. Table.\ref{Tab4} presents the quantitative comparison results of our method against 14 other methods. Compared to the state-of-the-art TransAttUnet-R \citep{ref40}, our PL-Net achieves better overall segmentation performance, with improvements ranging from 0.27\% to 1.16\% for different evaluation metrics. The segmentation performance of U-Net++ falls between our PL-Net† and PL-Net, with an IoU of 85.56\% and a Dice of 91.59\%. Across five cross-validation experiments, standard PL-Net showed higher stability than PL-Net†, with a 14\% reduction in standard deviation. Although the Dice score of Att R2U-Net is higher than that of PL-Net, its overall performance and stability are slightly inferior. Notably, both PL-Net and BiO-Net use BSC, but our method shows better overall performance. With a smaller PL-Net† size, almost the same IoU and Dice scores as BiO-Net (t = 3, INT) can be achieved.

\textbf{TN-SCUI and ACDC datasets.} The boundary of the TN-SCUI dataset is blurred compared to other datasets, and we found that methods including CNN may obtain better experimental results in this case. As shown in Table. \ref{Tab5}, even lightweight approaches like PL-Net† can achieve performance similar to Swin-UNet. UNeXt-L, a hybrid network based on CNN and MLP, has the smallest model size, but its segmentation performance is inferior to baseline methods. Our analysis shows that this is because the method has fewer learnable parameters and cannot make good use of the learned features. In the ACDC dataset (Table. \ref{Tab6}), we demonstrate the segmentation performance of different methods on different classes. The target area of the myocardium (Myo) is ringed between the left atrium (LV) and right atrium (RV) and is relatively small overall. The segmentation accuracy of different methods on this category tends to be lower than that of the other two categories. Our PL-Net achieves the highest IoU and Dice scores. Although TransUNet and EANet can achieve better average segmentation performance, their model size is increased by 6 times, making them more complex and requiring more computing resources than our proposed method. Additionally, the experimental results of PL-Net on the ACDC dataset show that our method is also suitable for multi-category segmentation tasks.

\begin{figure}
    \centering
    \includegraphics[width=0.95\textwidth]{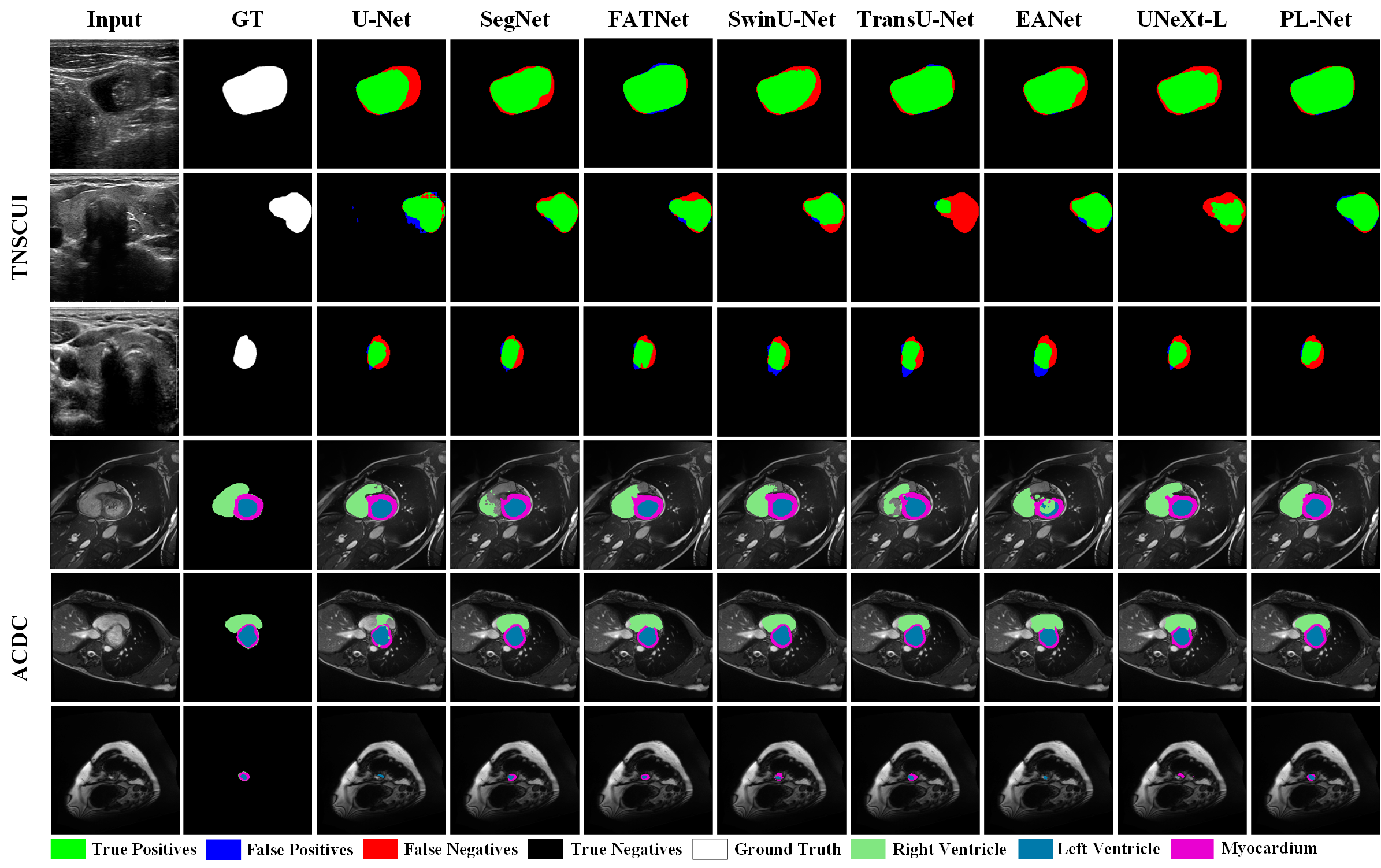}
    \caption{Qualitative segmentation results of TN-SCUI and ACDC datasets using different methods.}
    \label{fig5}
\end{figure}

The above quantitative comparison demonstrates that our proposed network can be applied to different segmentation tasks, which can include different modalities and categories. Even for images with blurred boundaries, PL-Net can produce good segmentation results. Although the overall segmentation performance of PL-Net† is not as good as that of standard PL-Net, its smaller parameters and model size will promote its application in memory-constrained environments. Additionally, other U-Net variants, which are oriented towards functional optimization or structural optimization, can improve the segmentation performance of the original U-Net to some extent, but the increased computational cost is a difficult problem to avoid. As PL-Net is a progressive learning framework, it achieves a good trade-off between segmentation performance and parameters.

\subsubsection{Qualitative Comparison}
To better understand the excellent performance of our method, we present example results of PL-Net and several other methods in Fig. \ref{fig4} and Fig. \ref{fig5}. As shown, our PL-Net and PL-Net† can handle different types of targets and produce accurate segmentation results.

The first and second rows of Fig. \ref{fig4} respectively show the segmentation results of an ambiguous target area and a small amount of occlusion (hair). As observed, although the results produced by PL-Net are not as accurate, our method is still effective for areas with ambiguous targets compared to other methods. When segmenting occluded images, other models either tend to divide boundaries incorrectly or mistake masked areas as target areas. The segmentation target of the image in the third row is clear, and relatively accurate segmentation results can be produced through other methods. However, for the content marked in the red box, most methods mistake interfering pixels for target pixels for segmentation, and better results are produced through our method compared to other methods. The fourth row shows the performance of different models for targets consisting of tiny targets and dispersed structures. As observed, U-Net and Att U-Net either recognize the saliency area as the target area or lose the target area, resulting in poor segmentation results. The fifth and sixth rows show the segmentation results of different methods for smaller and larger targets. As seen, our model makes a good decision on the boundary of the small target, while the area marked in the red box cannot be segmented well by other models. Compared to the fifth row, the lesion area shown in the sixth row covers almost the entire image. Although more accurate segmentation results can be produced through other methods, our PL-Net produces more perfect results as far as the area marked in the red box is concerned.

Fig. \ref{fig5} presents qualitative comparison results of different methods on the TN-SCUI and ACDC datasets. From the experimental results in the first two rows of the TN-SCUI dataset, PL-Net has a larger true positive area compared to other methods and is more accurate in lesion boundary segmentation. The third row shows an example where different methods perform poorly. Although there is a certain difference between our segmentation results and the ground truth, the false positive area is significantly lower than that of other methods, which is particularly important in medical image analysis. We highlighted different targets in the ACDC dataset using different colors, and the experimental results in the fourth-row show that SegNet, TransUNet, and EANet have poor segmentation results and incomplete segmentation of the RV area. In the example image in the fifth row, the Myocardium area accounts for a relatively small proportion, and FATNet, EANet, and UNeXt do not correctly segment the ring area, while PL-Net clearly segments the Myocardium area. The experimental results in the last row demonstrate the advantage of PL-Net in segmenting small targets. Although U-Net, EANet, and UNeXt segment the target area, their category definitions are inaccurate. These experimental results cover different situations in medical images, including large, medium, and small lesions, as well as targets of different categories. These results indicate that PL-Net has good generalization ability and can handle different types of medical image semantic segmentation problems.

\begin{figure}
    \centering
    \includegraphics[width=0.80\textwidth]{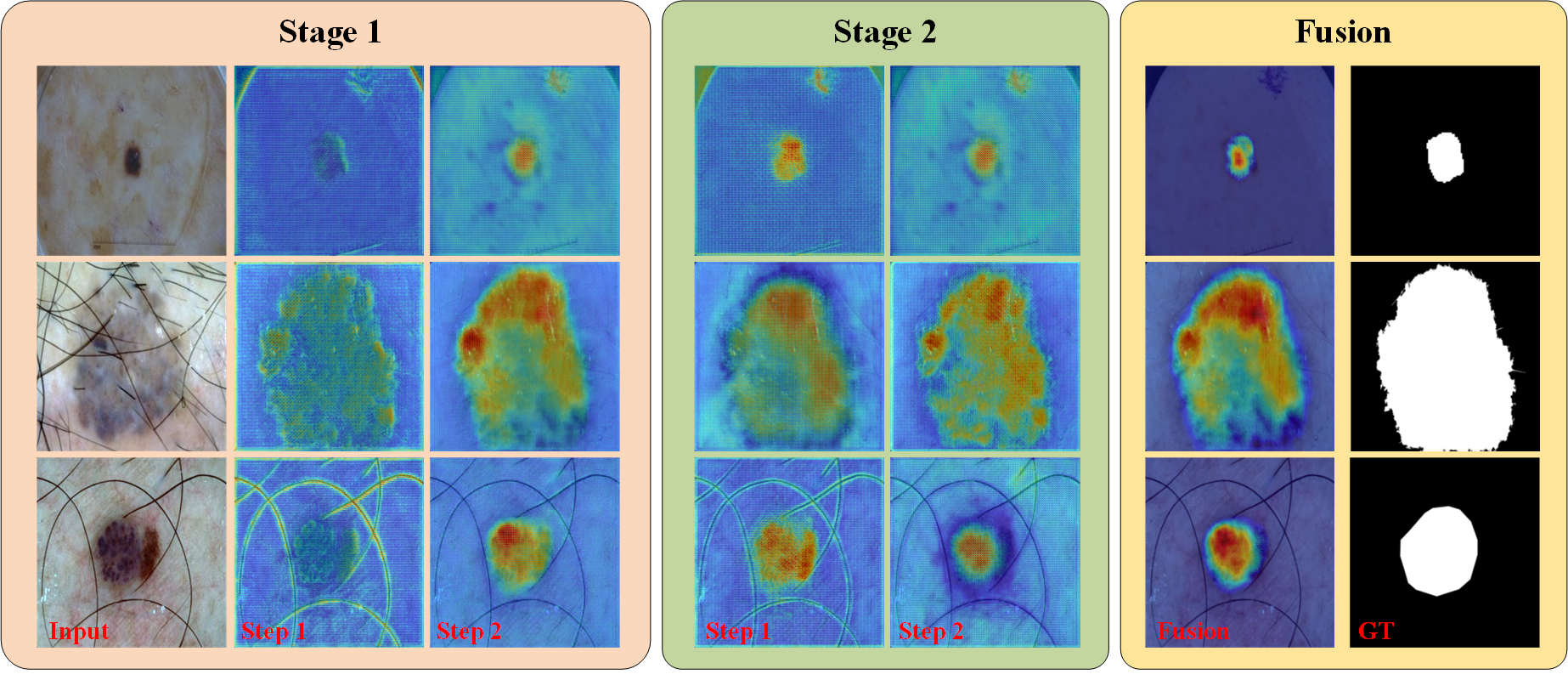}
    \caption{PL-Net's heat map of different "stages" and "steps" on the ISIC2017 dataset.}
    \label{fig6}
\end{figure}

In addition to the visualization results mentioned above, we present the features learned by different "stages" and "steps" of PL-Net in the form of a heat map, as shown in Fig. \ref{fig6}. During internal progressive learning (i.e., "Step1" and "Step2"), the shallower "Step1" tends to focus on coarse-grained semantic information first, such as the outline of hair or lesions. As the network depth increases, "Step2" gives less weight to texture features and focuses more on fine-grained semantics. PL-Net captures semantic information from coarse to fine granularity at different "stages" using internal progressive learning and does not introduce additional parameters compared to other approaches that replace deeper encoders. Through the visualization results of different "stages," we observe that the heat value of "Stage2" is higher than that of "Stage1" at the same position (i.e., the corresponding weight value is larger), which benefits from the fusion of coarse-grained and fine-grained information of the two stages. In addition, "Fusion" represents the feature map of the last convolutional layer after the two-stage fusion, with a distribution of thermal values similar to that of the ground truth and masked from irrelevant background regions.

\subsubsection{Expanding to 3D medical image segmentation}
%在本节中，我们将详细介绍如何将本文提出的渐进式策略有效地应用于3D医学图像分割任务中。为了验证该策略的有效性，我们选择了3D U-Net作为基线网络，并在前列腺磁共振成像（MRI）分割的标准数据集PROMISE 2012上进行了初步的实验验证。该数据集包含了50个MRI体积数据，我们按照3:2的比例划分为训练集和测试集。

%在3D U-Net的网络结构中，采用由两个3×3×3的三维卷积组成的基本块（此处省略了批量归一化和激活函数的描述）。编码器部分包含四个这样的基本块，每个基本块之后均进行了下采样操作，以逐步减少特征图的空间维度。解码器部分则通过四个上采样操作来恢复特征空间，每个上采样操作之后同样连接了一个基本块。

%为了将渐进学习策略融入3D U-Net中，我们将基本块中的两个3×3×3卷积转换为内部渐进学习过程，其中每个卷积层都视为一个“步骤”。在第二个“步骤”中，我们引入了向后跳跃连接，以在同一尺度上融合不同层次的特征。这种设计使得我们能够在不改变原始3D U-Net基本结构的前提下，有效地融入内部渐进学习策略。

%接下来，我们将上述网络视为外部渐进学习的首个“阶段”。为了构建第二个“阶段”，我们在编码器的末尾增加了一个下采样层和一个基本块，同时在解码器的开始处增加了一个上采样层和一个基本块。与PL-Net的设计理念类似，我们通过复用第一阶段的网络结构和新增的基本块来构建第二个“阶段”。在第二个阶段中，我们利用跳过连接将第一阶段的粗粒度信息与第二阶段的细粒度特征进行有效融合。通过上述步骤，我们将原始的3D U-Net扩展为一个具有两个“步骤”和两个“阶段”的渐进式学习网络。

%如表7所示，我们对比了原始3D U-Net与引入渐进式学习策略后的实验结果。从数据中可以看出，仅引入内部渐进学习策略就能使Dice分数相较于基线3D U-Net提升0.6%。而当同时应用两种渐进学习策略时，Dice分数的提升更为显著，达到了1.36%。这些初步的实验结果充分证明了我们所提出方法的有效性和实用性。受这些积极发现的鼓舞，我们计划在未来进一步探索渐进学习策略在更广泛的3D医学图像分割任务中的潜在性能。

{In this section, we will detail how to effectively apply the proposed progressive strategy to 3D medical image segmentation tasks. To validate the effectiveness of this strategy, we chose 3D U-Net as the baseline network and conducted preliminary experimental validation on the standard prostate MRI segmentation dataset, PROMISE 2012~\cite{litjens2014evaluation}. This dataset contains 50 MRI volumes, which we split into a training set and a test set in a 3:2 ratio.}

{In the 3D U-Net~\cite{3DU-Net} structure, we employed basic blocks consisting of two 3×3×3 convolutions (excluding batch normalization and activation functions here). The encoder part includes four such basic blocks, each followed by a downsampling operation to progressively reduce the spatial dimensions of the feature maps. The decoder part restores the feature space through four upsampling operations, each also followed by a basic block.}

{To integrate the progressive learning strategy into the 3D U-Net, we converted the two 3×3×3 convolutions in the basic block into an internal progressive learning process, where each convolution layer is considered a "step." In the second "step," we introduced backward skip connections to fuse features of different levels at the same scale. This design allows us to effectively incorporate the internal progressive learning strategy without altering the original 3D U-Net's basic structure.}

{Next, we regarded the above network as the first "stage" of external progressive learning. To construct the second "stage," we added a downsampling layer and a basic block at the end of the encoder, and an upsampling layer and a basic block at the beginning of the decoder. Similar to the design concept of PL-Net, we built the second "stage" by reusing the network structure from the first stage along with the newly added basic blocks. In the second stage, we used skip connections to effectively fuse the coarse-grained information from the first stage with the fine-grained features of the second stage. Through these steps, we extended the original 3D U-Net into a progressive learning network with two "steps" and two "stages."}

{As shown in Table 7, we compared the experimental results of the original 3D U-Net with those after introducing the progressive learning strategy. The data indicates that introducing only the internal progressive learning strategy improved the Dice score by 0.6\% compared to the baseline 3D U-Net. When both progressive learning strategies were applied, the Dice score improvement was even more significant, reaching 1.36\%. These preliminary experimental results fully demonstrate the effectiveness and practicality of our proposed method. Encouraged by these positive findings, we plan to further explore the potential performance of progressive learning strategies in a broader range of 3D medical image segmentation tasks in the future.}

\begin{table}[ht]\small\centering
\caption{{Extended experiments on the application of progressive learning strategies to 3D U-Net.}}
\begin{tabular}{l|cc}
\hline
\multirow{2}{*}{Method} & \multicolumn{2}{c}{PROMISE 2012 Dataset} \\ \cline{2-3} 
                        & IoU                 & Dice               \\ \hline
3D U-Net~\cite{3DU-Net}                & 56.84\%             & 72.48\%            \\
3D U-Net+IPL            & 57.58\%             & 73.08\%            \\
3D U-Net+IPL+EPL        & \textbf{58.53\%}             & \textbf{73.84\%}            \\ \hline
\end{tabular}
\end{table}

\section{Discussion}

U-Net has been widely used as a benchmark model for medical image segmentation due to its simple and easily modifiable structure. Most of its variant approaches enhance segmentation performance by adding functional modules (e.g., attention module) or modifying its original structure (e.g., residual, and densely connected structures) in the feed-forward process. In this paper, we adopt an alternative approach by recognizing that coarse-grained and fine-grained discriminative information naturally exists at different stages of the network, which can be learned incrementally, similar to how humans learn through shallow and deep network structures. Based on this intuition, we design a framework with internal and external progressive learning strategies, called PL-Net. Internal progressive learning strategies are used to mine semantic information at different granularities, while external progressive learning strategies further refine segmentation details based on the features learned in the previous training phase.

Researchers have proposed numerous network architectures based on U-Net to address various medical image semantic segmentation problems. However, some approaches that add functional modules (such as FCANet and Att U-Net) do not consistently improve performance across different datasets. Our experimental results demonstrate that while FCANet improves IoU by 4\% over vanilla U-Net on the ISIC2017 dataset, it degrades performance by 0.81\% on the Nuclei dataset, indicating that performance variation is related to the type, size, and complexity of the dataset. Our proposed PL-Net achieves consistent performance improvements over vanilla U-Net on five datasets without adding new functional modules or structural modifications and remains competitive with state-of-the-art network frameworks (EANet and ResGANet). Moreover, PL-Net has lower computational overhead and fewer parameters, resulting in a model size reduction of 3.8 times and 6.6 times compared to widely used nnUNet and TransUNet, respectively. We also provide PL-Net† with a smaller number of parameters, which can offer options for different medical imaging scenarios, although the decrease in the number of parameters results in reduced segmentation accuracy. Our method can run on a GPU with limited memory, reducing the complex configuration and tedious preprocessing steps of nnUNet. In other words, designing such a network is crucial to translate medical imaging from the laboratory to clinical practice.

On the other hand, similar to most existing state-of-the-art methods, our proposed segmentation network still has limitations in handling cases with complex boundaries and small targets. As shown in the first row of Fig. \ref{fig4}, when the boundary between the skin lesion and the background region is difficult to distinguish, our method and other approaches fail to accurately delineate the boundary. As shown in the third row of Fig. \ref{fig5}, PL-Net's segmentation performance is lower when the target region is very small. However, in these cases, our method is closest to the ground truth, and the segmentation results are still better than those of other competitors. From the experimental results in Table.\ref{Tab1}, we found that the best results were obtained by performing three internal progressive learning experiments on the large-scale TN-SCUI dataset, indicating the necessity of setting different internal progressive learning strategies. Finally, we believe that introducing robust functional modules may further improve the segmentation performance of PL-Net, and we will explore this in future work. The ideas proposed in this paper mainly provide inspiration for researchers who are committed to designing feature representations to improve convolutional neural networks.

\section{Conclusion}

In this study, we propose a new variant of U-Net called PL-Net for 2D medical image segmentation, which mainly consists of internal and external progressive learning strategies. Compared to U-Net methods that optimize functional or structural aspects, our PL-Net achieves consistent performance improvements without additional trainable parameters. We provide both a standard PL-Net (15.03 M) and a smaller version, PL-Net† (3.77 M), to address different medical image segmentation scenarios in real-world situations. We conduct comprehensive experiments on five public medical image datasets, and the results show that PL-Net can improve the segmentation IoU of the baseline network by 0.46\% to 4.9\%, demonstrating high competitiveness with other state-of-the-art methods.

{Although our proposed method has shown promising results, it still has some limitations that need to be further addressed in future research: 1) Impact of data size: Exploring the parameter settings of internal and external progressive learning under different data sizes will help researchers understand the potential of the model under different scales of data. In the future, we will further explore the performance of PL-Net on larger datasets. 2) Due to the limitations of computing power and data, our method mainly focuses on 2D medical image segmentation. This article has initially demonstrated the feasibility of the progressive learning strategy in 3D medical image segmentation. In the future, we will extend PL-Net to more advanced 3D medical image segmentation frameworks to further enhance its capabilities in 3D medical image segmentation. 3) Design of functional modules: How to design functional modules suitable for PL-Net to improve segmentation performance while maintaining a concise framework is also a topic for further research in the future.}

%In future research, we plan to investigate more robust medical image segmentation models and explore the performance of this framework in three-dimensional medical image segmentation. Additionally, as PL-Net is a flexible architecture, we should pay more attention to its ability to generalize to large-scale medical image data, providing help for research on general medical image segmentation.

\section*{Author Contributions}
K.Mao: Writing – review \& editing, Conceptualization, Methodology; R.Li: Writing – review \& editing, Data processing, Visualization; J.Cheng: Writing – original draft, Conceptualization, Methodology; D.Huang: Writing – original draft, Formal Analysis, Experiment; Z.Song: Experiment, Visualization; Z.Liu: Data processing, Methodology, Project administration.

\section*{Conflict of interest}
The authors declare that the research was conducted in the absence of any commercial or financial relationships that could be construed as a potential conflict of interest.
%The Author Contributions section is mandatory for all articles, including articles by sole authors. If an appropriate statement is not provided on submission, a standard one will be inserted during the production process. The Author Contributions statement must describe the contributions of individual authors referred to by their initials and, in doing so, all authors agree to be accountable for the content of the work. Please see  \href{https://www.frontiersin.org/about/policies-and-publication-ethics#AuthorshipAuthorResponsibilities}{here} for full authorship criteria.

\section*{Funding}
This study was partially funded by Chongqing Municipal Education Commission Science and Technology Research Project (KJQN202203302).

%\section*{Supplemental Data}
 %\href{http://home.frontiersin.org/about/author-guidelines#SupplementaryMaterial}{Supplementary Material} should be uploaded separately on submission, if there are Supplementary Figures, please include the caption in the same file as the figure. LaTeX Supplementary Material templates can be found in the Frontiers LaTeX folder.

\section*{Data Availability Statement}
The raw data supporting the conclusions of this article will be made available by the authors, without undue reservation.
%The datasets [GENERATED/ANALYZED] for this study can be found in the [NAME OF REPOSITORY] [LINK].
% Please see the availability of data guidelines for more information, at https://www.frontiersin.org/about/author-guidelines#AvailabilityofData

\bibliographystyle{Frontiers-Harvard} %  Many Frontiers journals use the Harvard referencing system (Author-date), to find the style and resources for the journal you are submitting to: https://zendesk.frontiersin.org/hc/en-us/articles/360017860337-Frontiers-Reference-Styles-by-Journal. For Humanities and Social Sciences articles please include page numbers in the in-text citations 

\bibliographystyle{unsrt}

%%\bibliography{bib_file}
%%% Make sure to upload the bib file along with the tex file and PDF
%%% Please see the test.bib file for some examples of references

%\section*{Figure captions}

%%% Please be aware that for original research articles we only permit a combined number of 15 figures and tables, one figure with multiple subfigures will count as only one figure.
%%% Use this if adding the figures directly in the mansucript, if so, please remember to also upload the files when submitting your article
%%% There is no need for adding the file termination, as long as you indicate where the file is saved. In the examples below the files (logo1.eps and logos.eps) are in the Frontiers LaTeX folder
%%% If using *.tif files convert them to .jpg or .png
%%%  NB logo1.eps is required in the path in order to correctly compile front page header %%%

\end{document}